\documentclass[aps,prb,amsmath,twocolumn]{revtex4}

\usepackage{graphicx}
\usepackage{bm}

\begin{document}

\title{Current-biased Andreev interferometer}

\author{Artem V. Galaktionov$^{1,2}$
and Andrei D. Zaikin$^{3,1}$
}
\affiliation{$^1$I.E.Tamm Department of Theoretical Physics, P.N.Lebedev
Physical Institute, 119991 Moscow, Russia\\
$^2$Laboratory of Cryogenic Nanoelectronics, Nizhny Novgorod State Technical University, 603950 Nizhny Novgorod, Russia\\
$^3$Institute of Nanotechnology, Karlsruhe Institute of Technology (KIT), 76021 Karlsruhe, Germany
}

\begin{abstract}
We theoretically investigate the behavior of Andreev interferometers with three superconducting electrodes in the current-biased regime. Our analysis allows to predict a number of interesting features of such devices, such as both hysteretic and non-hysteretic behavior, negative magnetoresistance and two different sets of singularities of the differential resistance at subgap voltages. In the non-hysteretic regime we find a pronounced voltage modulation with the magnetic flux which can be used for improving sensitivity of Andreev interferometers.

\end{abstract}

\maketitle

\section{Introduction}

It was demonstrated experimentally by Petrashov and co-workers \cite{Petr1} that the resistance of mesoscopic diffusive normal-superconducting (NS) hybrid structures can be quite strongly modulated by an externally applied magnetic field. These observations open a possibility to construct
the so-called Andreev interferometers which -- for a number of applications -- may have several important advantages over the well known Superconducting Quantum Interference Devices (SQUIDs). Such kind of applications include, e.g., read-out of superconducting qubits\cite{Petrreadout} or experimental analysis of switching dynamics of individual magnetic nanoparticles\cite{Petrdynam} and require such features of a detecting device as reduced intrinsic dissipation (down to fW level), the possibility to achieve higher sensitivity and read-out speed as well a broad choice of normal conductors employed as a weak link.

Recently, there were both experimental \cite{MM} and theo\-retical \cite{GT,GZK} investigations of hybrid NS structures with pronounced dependencies of their resistance on the applied magnetic flux. Contrary to the initial design of Andreev interferometers pioneered by Petrashov and co-authors \cite{Petr1}, this latest analysis focuses on systems with all external electrodes in the superconducting state.  In this way one
would be able to reduce dissipation.  Current (voltage) harmonics
would be generated in this case, which are multiples of the Josephson
frequency. However, they can be filtered out and the average current (voltage) can be measured which should reveal a dependence on the magnetic flux.

In Ref. \onlinecite{GZK} we already carried out a detailed theoretical analysis
of Andreev interferometers in the voltage-biased regime. While this regime
can be realized in some experiments, of a clear experimental interest is also another physical situation when the system is biased by a fixed external current.  The main purpose of the present work is to analyze
the behavior of Andreev interferometers in the current-biased regime.

Note that the system behavior in the latter regime can be very different
from that in the voltage-biased one. In a vast majority of normal structures these differences mainly concern higher cumulants of voltage and current. For instance, a decade ago there arose a conundrum caused by experiments \cite{Reul}, where the third voltage cumulant of the current-biased normal junction was measured. The behavior of this quantity was essentially different from  theoretical expectations based on the linear relation between the third voltage correlator in the current-biased regime and the third current correlator \cite{Levitov2,3rdcum,Salo} in the voltage-biased regime. This conundrum was resolved in Ref. \onlinecite{Kind} which major conclusion was that current and voltage correlators of order three and higher are no longer linearly related.

In superconducting circuits essential differences between the voltage- and current-biased regimes occur already at the level of $I-V$ curves. Such
differences were encountered, e.g., in the case of single Josephson junctions \cite{McD} (see also Ref. \onlinecite{Likh}).
The experiments \cite{Niem} conformed to the corresponding theoretical predictions. Another example is the low temperature behavior of Josephson
junction arrays and chains which may vary from superconducting to insulating depending on whether the voltage- or current-biased scheme is considered \cite{BFSZ}.

The structure of our paper is as follows. In Sec. II we describe the
model for Andreev interferometers under consideration. In Sec. III we proceed with theoretical analysis of our model and derive the basic
formula that fixes the current-phase relation for our device. This formula is then employed in Sec. IV where we numerically evaluate the $I-V$ curve
for our system in the current-biased regime and study the effect of voltage modulation by an external magnetic flux. The paper is concluded by a brief discussion of our key observations in Sec. V. Some technical details are
relegated to Appendix.

\section{The Model}

\begin{figure}
\includegraphics[width=7cm]{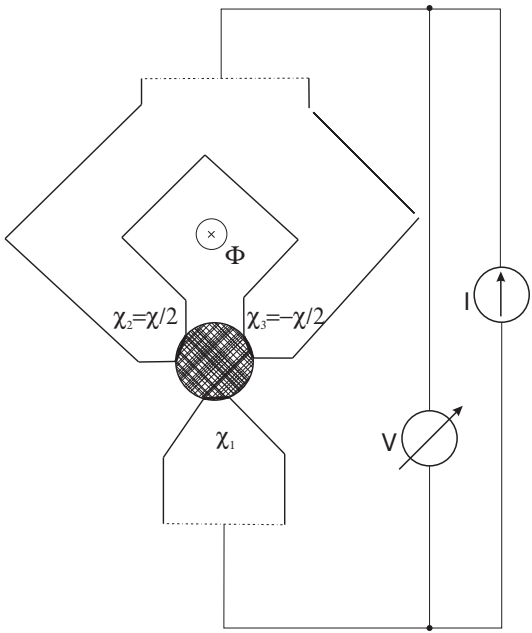}
\caption{Scheme of the setup: disordered normal metal insertion ("dot") embedded between three superconducting electrodes. The phase difference $\chi_2-\chi_3=\chi$ is
 caused by the magnetic flux $\Phi$ piercing the loop.}
\label{setup}
\end{figure}

The system under consideration is schematically de\-picted in Fig. \ref{setup}.
It consists of three superconducting electrodes characterized by the absolute value of the order parameter $\Delta$
and a disordered normal metal insertion embedded between these electrodes.
A typical size $L$ of this normal metallic dot is assumed not to exceed the superconducting coherence length $\xi$
and at the same time to be larger than the elastic mean free path $\ell$. Two superconducting electrodes 2 and 3 form a loop which
is pierced by an external magnetic flux $\Phi$. Accordingly, the superconducting phase difference $\chi =2\pi \Phi/ \Phi_0$ is
induced between the electrodes 2 and 3, where $\Phi_0$ is the superconducting flux quantum.

In what follows we will generally assume that all interfaces between the normal metal and superconducting electrodes are weakly transmitting. In this case one can expect to observe a pronounced magnetoresistance modulation effect \cite{GWZ}. For comparison, in the limit of highly transparent NS interfaces this modulation is expected to remain below ten percent \cite{NazSt,GWZ}. The normal state conductances $G_1$, $G_2$ and $G_3$ between three electrodes and the normal insertion obey the condition
\begin{equation}
 G_{1} \equiv \frac{1}{R_N}\ll G_{2,3} \ll \sigma_D {\cal A}/L.
\label{tunlim}
\end{equation}
where the normal state resistance $R_N$ is determined by the standard Landauer formula
\begin{equation}
\frac{1}{R_N}=\frac{e^2}{\pi} \sum_n T_n,
\end{equation}
$T_n$ are the transmissions of conducting channels in the first barrier,
$\sigma_D=2e^2 DN_0$ is the Drude conductivity of the normal metal, ${\cal A}$ denotes a typical contact area between the normal metal
and the electrode, $D=v_F\ell /3$ is
the diffusion coefficient and $N_0$ is the density of states at the Fermi-surface per spin
direction. The electron charge will be denoted by $-e$.

Eq. (\ref{tunlim}) assures that the voltage
drop occurs only across the tunnel barrier between the first electrode and the rest of our system. The second inequality (\ref{tunlim}) also guarantees that our results will not depend on the particular shape of the normal metal insertion.
In addition, we will disregard charging effects which amounts to assuming that all relevant effective charging energies remain much smaller
than other important energy scales in our problem \cite{SZ}.

As we already discussed, below we will be interested in the current-biased regime, i.e. we will
assume that our system is biased by an external current source $I$, as it is indicated in Fig. \ref{setup}.
Restricting ourselves to this regime we will evaluate the time-averaged voltage $V$ across our device which should depend both on
the bias current $I$ and on the external flux $\Phi$ (or, equivalently, on the phase difference $\chi$). In the case of Josephson
junctions with normal state resistance $R_N$ and capacitance $C$ it was demonstrated \cite{McD} that the $I-V$ curves in the current- and
voltage-biased regimes differ substantially provided the parameter $\beta=\Delta R_N C$ remains smaller than one. Below we will also
stick to the same limit $\beta \ll 1$.

\section{Current-phase relations}

The first step to derive the current-voltage characteristics is to establish the dependence of the instantaneous current value $I(t)$ on the phase difference $\varphi(t)$ across the NS interface with the smallest conductance $1/R_N$. This phase difference is defined by the standard relation
\begin{equation}
\varphi(t)=e\int\limits_0^t dt' V(t')
\end{equation}
where $V(t)$ is the time dependent voltage drop across this NS interface.
In order to accomplish this goal one can employ the effective action analysis \cite{SZ,Z94,Sny}. The general Keldysh effective action describing electron transport across the barrier between the first superconducting electrode and the rest of our structure has the form
\begin{equation}
{\cal S}=-\frac{i}{2}\sum_n{\rm Tr} \ln \left[ 1+\frac{T_n}{4}\left( \left\{ \check
g_N, \check g_S\right\}-2\right) \right].\label{pact}
\end{equation}
Here $\check g_{S}$ is the Green-Keldysh matrix of the superconducting reservoir 1 and $\check g_{N}$ is the $4\times 4$ Green-Keldysh matrix of a dis\-ordered normal metal which also acquires superconducting properties due to the contact with the reservoirs 2 and 3.

The effective action (\ref{pact}) holds for arbitrary transmission values $T_n$ and allows for a complete description of electron transport through our device. It is convenient to combine the phase variables on the two branches of the Keldysh contour and define the "classical" ($\varphi (t)$) and "quantum" ($\varphi_-(t)$) components of the phase. Taking the derivatives of the effective action with respect to the quantum phase $\varphi_-$ one can derive the expressions for all current correlators in our problem. Provided charging effects are weak the action (\ref{pact}) can be expanded in powers of $\varphi_-$ and reduced to a much simpler form convenient for practical calculations. In the case of NS hybrid structures this procedure was described in details in Ref. \onlinecite{GZ10}.

Since here we are merely interested in the tunneling limit $T_n \ll 1$, it suffices to expand the action (\ref{pact}) up to the first order in $T_n$.
Then taking the first variation of the action with respect to $\varphi_-$ one arrives at the current-phase relation in the form
\begin{eqnarray}
&& I(t)=\int\limits_0^t dt'\sin\left[ \varphi(t) -\varphi(t')\right]S_1(t-t')\nonumber\\ && +\int\limits_0^t dt'\sin\left[ \varphi(t) +\varphi(t')\right]S_2(t-t'), \label{cphr}
\end{eqnarray}

The kernels $S_{1}(t)$ and $S_{2}(t)$ are expressed respectively via normal and anomalous components of the Green-Keldysh matrices, i.e.
\begin{eqnarray}
&& S_1(t)=-\frac{ie}{4}\sum_n T_n \left[ g_S^R(t)g_N^K(-t)+ g_S^K(t)g_N^A(-t)\right.\nonumber\\&& \left. +g_N^R(t)g_S^K(-t)+ g_N^K(t)g_S^A(-t)\right],\label{gr}\\
&& S_2(t)=\frac{ie}{4}\sum_n T_n \left[ f_S^R(t)f_N^K(-t)+  f_S^K(t)f_N^A(-t)\right.\nonumber\\ &&\left. +f_N^R(t)f_S^K(-t)+ f_N^K(t)f_S^A(-t)\right]. \nonumber
\end{eqnarray}

Note that Eqs. (\ref{cphr}), (\ref{gr}) can also be derived by means of the standard tunneling Hamiltonian approach.

What remains is to define the Green functions of both the superconducting electrode and the normal metal dot. Without any loss of generality one can set the electric potential of the first superconducting electrode equal to zero. Then
the Fourier transforms of $g_S^{R,A}$ and $f_S^{R,A}$ take the form
\begin{equation}
g_S^{R,A}(\epsilon)=\frac{\epsilon}{\xi^{R,A}}, \quad f_S^{R,A}(\epsilon)=\frac{\Delta}{\xi^{R,A}}.
\end{equation}
In order to properly account for the analytic properties of the functions $\xi^{R,A}$ it is important to keep an infinitesimally small imaginary part $i0$, i.e.
\begin{equation}
\xi^{R,A}=\pm\sqrt{(\epsilon\pm i0)^2-\Delta^2}.
\end{equation}
As the cut in the complex plane goes from $-\Delta$ to $\Delta$, we obtain
 $\xi^{R,A}=\pm{\rm sgn}\,\epsilon\sqrt{\epsilon^2-\Delta^2}$ for $|\epsilon|>\Delta$ and  $\xi^{R,A}=i\sqrt{\Delta^2-\epsilon^2}$  for  $|\epsilon|<\Delta$.

 The Keldysh components $g^K$ and $f^K$ are related to the above retarded and advanced reen functions in the standard manner as
 \begin{eqnarray}
&& g^K(\epsilon)=\left(g^{R}(\epsilon)-g^A(\epsilon) \right)\tanh\frac{\epsilon}{2T},\label{KKD}\\ && f^K(\epsilon)=\left(f^{R}(\epsilon)-f^A(\epsilon) \right)\tanh\frac{\epsilon}{2T}.\nonumber
 \end{eqnarray}

Now let us turn to the Green-Keldysh functions of the metallic dot $g_N^{R,A,K}$ and $f_N^{R,A}$. These functions have already been evaluated elsewhere \cite{BSW,GZK}, therefore here we only briefly recapitulate the corresponding results. It is important to bear in mind that due to the contact with superconducting terminals 2 and 3 the normal metal also acquires superconducting properties. For instance, the proximity-induced minigap $\Delta_g$ in its spectrum develops. This minigap is defined by the equation \cite{GZK}
\begin{equation}
 \Delta_g=\frac{\epsilon_g}{1+ \gamma\sqrt{1-\Delta_g^2/\Delta^2}},\label{mge}
 \end{equation}
where the quantity
\begin{equation}
\epsilon_g=\Delta\sqrt{1-\frac{4 G_2 G_3}{(G_2+G_3)^2}\sin^2\frac{\chi}{2}}
\label{bex}
\end{equation}
depends on the external magnetic flux $\Phi$ via the phase difference $\chi$. The parameter $\gamma$ \cite{BSW} effectively controls the strength of electron-hole dephasing in our system.  This parameter is defined as
 \begin{equation}
\gamma=\frac{2\sigma_D {\cal V}\Delta}{D(G_2+G_3)}, \label{gdef}
\end{equation}
where ${\cal V}$ stands for the volume of the normal metal.

In order to correctly determine analytic properties of the Green functions we observe that the structure of the cuts in the complex plane is somewhat more complicated. Namely these cuts are now located at $(-\infty,-\Delta]$, $[-\Delta_g,\Delta_g]$ and $[\Delta,\infty)$. As above, the retarded Green functions are defined on the upper banks of these cuts. Provided $\Delta_g<\epsilon<\Delta$ we find
\begin{eqnarray}
&& g_N^R(\epsilon)=\frac{\epsilon}{\sqrt{\epsilon^2-\frac{\epsilon_g^2}{\left( 1+\gamma\sqrt{1-\epsilon^2/\Delta^2}\right)^2}}},\\
&& f_N^R(\epsilon)=\frac{\epsilon_g}{\sqrt{\epsilon^2\left( 1+\gamma\sqrt{1-\epsilon^2/\Delta^2}\right)^2-\epsilon_g^2}}.\nonumber
\end{eqnarray}
For the remaining values of $\epsilon$ the retarded Green functions are obtained by analytic continuation with the mentioned cuts, while advanced Green functions are defined as $g_N^A(\epsilon)=-\left[ g_N^R(\epsilon) \right]^*$ and $f_N^A(\epsilon)=-\left[ f_N^R(\epsilon) \right]^*$. Finally, the Keldysh components are again given by Eqs. (\ref{KKD}).

\begin{figure}
\includegraphics[width=8cm]{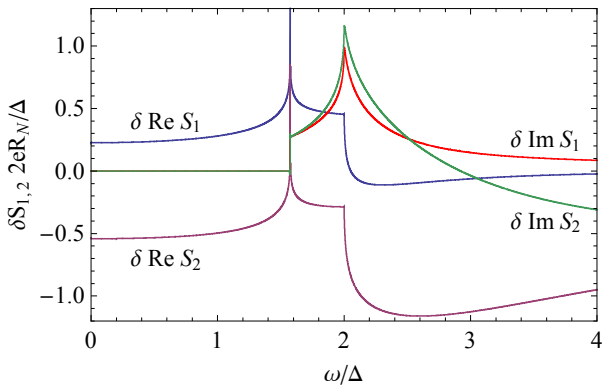}
\caption{The functions $\delta S_{1,2}(\omega)$ (\ref{ds}) evaluated at $T \to 0$ for $\epsilon_g=0.95$ and $\gamma=0.8$. }
\label{ztk}
\end{figure}

Combining the above expressions for the Green functions with Eqs. (\ref{gr}) it is easy to verify that $S_{1,2}(t<0)\equiv 0$, i.e. both kernels (\ref{gr}) obey the requirement of causality. In the case of a Josephson junction between two BCS superconductors with different values of the gap the kernels $S_{1,2}(\omega)$ were derived by Werthamer \cite{W} and also by Larkin and Ovchinnikov \cite{LO}. For reference purposes the corresponding expressions (denoted below as $\tilde S_{1,2}(\omega)$) are presented in Appendix.
In our case the kernels $S_{1,2}(\omega)$ deviate from $\tilde S_{1,2}(\omega)$
since the energy spectrum of the central metallic dot is different from that of a superconductor. The difference
\begin{equation}
\delta S_{1,2}(\omega)=S_{1,2}(\omega)-\tilde S_{1,2}(\Delta,\Delta_g,\omega)
\label{ds}
\end{equation}
can be evaluated both numerically \cite{avail} and analytically in some limits.
For illustration, in Fig. \ref{ztk} we display the functions $\delta S_{1,2}(\omega)$ calculated at $\epsilon_g=0.95$, $\gamma=0.8$ and $T \to 0$.

Within the logarithmic accuracy an asymptotic behavior  of ${\rm Re}\,\delta\, S_{1,2} (\omega)$ at $\omega\approx \Delta+\Delta_g$ is described as
\begin{equation}
{\rm Re}\,\delta\, S_{1,2} (\omega)\approx \frac{\Delta}{2eR_N}\sqrt{\frac{\Delta_g}{\Delta}}\left(\frac{1}{\sqrt{c_1}}-1\right) \ln\frac{\Delta_g}{|\delta\omega|},
\end{equation}
where $\delta\omega=\omega-\Delta-\Delta_g$ and
\begin{equation}
c_1=1-\frac{\gamma^2\Delta_g^4}{\Delta^2\epsilon_g(\epsilon_g-\Delta_g)}.
\end{equation}
These expressions imply modifications in the so-called Riedel singularity \cite{Ried} as compared to the case of usual Josephson junctions between two superconductors. This difference is by no means surprising since the density of states in our metallic dot differs from that of a BCS superconductor. We also observe that the functions $  {\rm Im}\, \delta\,S_{1,2} (\omega)$ experience a jump at $\delta\omega=0$. The magnitude of this jump reads
\begin{equation}
{\rm Im}\,\delta\, S_{1,2} =\frac{\Delta}{2eR_N}\left[\pi \sqrt{\frac{\Delta_g}{\Delta}} \left(\frac{1}{\sqrt{c_1}}-1\right) \right].
\end{equation}
Finally, we note the presence of peculiarities in the behavior of the functions $ {\rm Re}\,\delta\, S_{1,2}(\omega)$ and ${\rm Im}\,\delta\, S_{1,2}(\omega)$ at $\omega=2\Delta$, cf. Fig. \ref{ztk}.

\begin{figure}
\includegraphics[width=8cm]{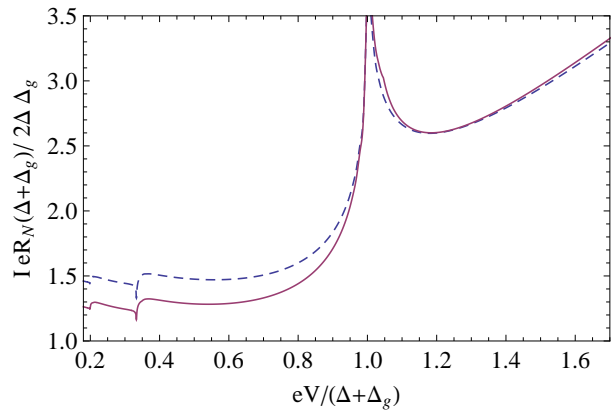}
\caption{The current through our device as a function of the average voltage at $T \to 0$ for $\epsilon_g=0.95$ and $\gamma=0.1$ (solid curve). The dashed curve corresponds to the same quantity evaluated with the kernels $\tilde S_{1,2}(\Delta, \Delta_g,\omega)$. }
\label{cv1}
\end{figure}

\section{Numerical analysis}

Our numerical procedure follows closely that of Ref. \onlinecite{McD}. We will make use of the representation
\begin{equation}
e^{i\varphi(t)}=e^{ieVt}\sum_{n=-N}^{N} W_n e^{-i n\omega_J t},\label{Wn}
\end{equation}
where $V$ is the average voltage across the tunnel barrier, $\omega_J=2eV$ is the Josephson frequency and $W_n$ are $2N+1$ complex numbers to be determined. As usually, Eq. (\ref{Wn}) demonstrates that harmonics with higher Josephson frequencies are excited in our device.

Our numerics shows that sufficient accuracy is achieved if one restricts the summation in Eq. (\ref{Wn}) by $N=25$. The numbers $W_n$ are determined bearing in mind that: \newline
({\it a}) In the current biased regime considered here only the $n=0$ component of the current differs from zero,\newline
({\it b}) The condition $e^{i\varphi(t)}e^{-i\varphi(t)}=1$ imposes extra restrictions on $W_n$ and \newline
{(\it c)} Here we choose $\varphi(t=0)=0$, which is equivalent to $\sum_{n=-N}^N {\rm Im } W_n=0$.

\begin{figure}
\includegraphics[width=7.8cm]{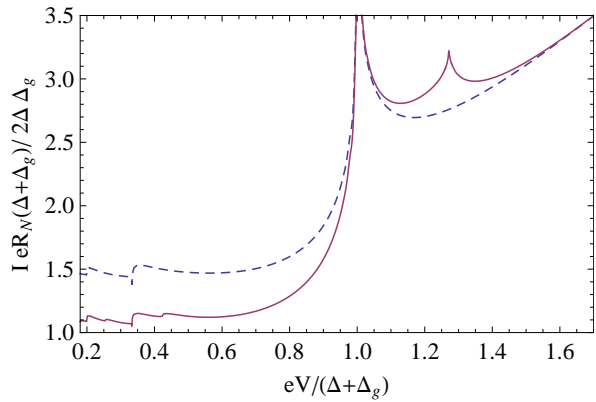}
\caption{ The same as in Fig. \ref{cv1} for $\epsilon_g=0.95$ and $\gamma=0.8$.}
\label{cv2}
\end{figure}

\begin{figure}
\includegraphics[width=7.8cm]{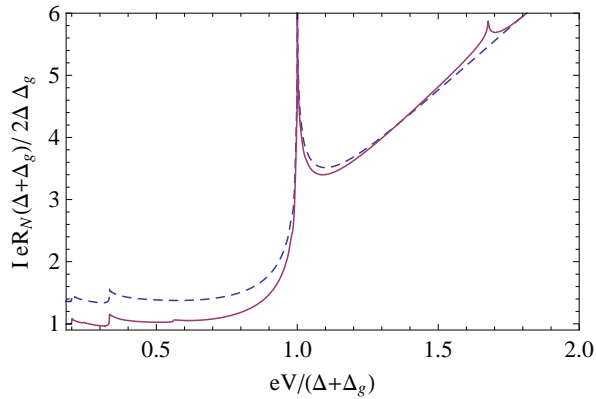}
\caption{ The same as in Fig. \ref{cv1} for $\epsilon_g=0.95$ and $\gamma=4$.}
\label{cv3}
\end{figure}

This set of conditions provides $4N+2$ real equations sufficient to fully determine $W_n$. Resolving these equations by the Newton's method we finally recover the current through our system as a function of the average voltage $V$. The results of numerical analysis are displayed in Figs. \ref{cv1}--\ref{cv3}.

Similarly to ordinary Josephson junctions between two BCS superconductors \cite{McD,Likh}  the $I-V$ curves demonstrate peculiarities at voltage values
\begin{equation}
eV=\frac{\Delta+\Delta_g}{2m+1},
\label{p1}
\end{equation}
where $m$ is an integer number. These peculiarities stem from the Riedel-like singularity contained in the kernels $S_{1,2}$. In addition, we also observe extra peculiarities which occur at voltages
\begin{equation}
eV=\frac{2\Delta}{2m+1},
\label{p2}
\end{equation}
These latter features are not present in ordinary Josephson junctions at all. In our case these peculiarities are caused by the behavior of kernels $S_{1,2}(\omega)$ at $\omega=2\Delta$, see Fig. \ref{ztk}. These additional features on the $I-V$ curve are more pronounced for intermediate values of the electron-hole dephasing parameter $\gamma\sim 1$ and become considerably less pronounced both at small and large values of $\gamma$, cf. Fig. \ref{cv2} with Figs. \ref{cv1} and \ref{cv3}.

\begin{figure}
\includegraphics[width=7.8cm]{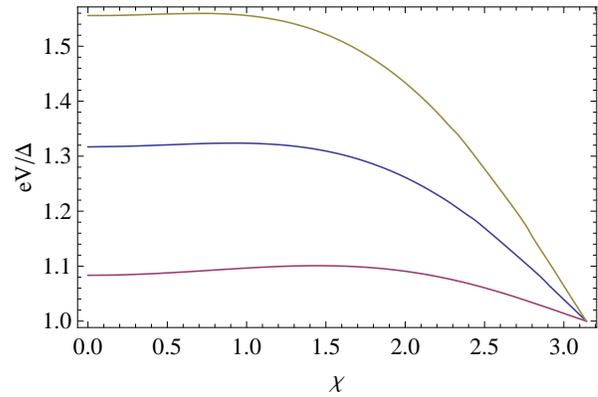}
\caption{Voltage modulation in a current-biased Andreev interferometer in the low temperature limit. Here we set $G_2=G_3$ and the parameter $\gamma$ equal to 0.1, 0.8, 4 (top to bottom). The corresponding current bias values are 1.4, 1., and 0.5 $\Delta/(e R_N)$.}
\label{mr1}
\end{figure}

The dependencies presented in Figs. \ref{cv1}--\ref{cv3} demonstrate that at certain values of the bias current the average voltage $V$ becomes multivalued, i.e. there
exist more that one different voltage states corresponding to the same bias $I$. Accordingly, in this regime one can
expect to observe jumps between different voltage branches as well a hysteretic behavior of our device. On the other hand, there also exists a subgap voltage regime where $V$ remains single valued for a fixed bias current. For example, for the parameters employed in Figs. \ref{cv1}-\ref{cv3} such non-hysteretic regime is realized within the voltage interval
\begin{equation}
 0.8\lesssim \frac{eV}{\Delta+\Delta_g} \lesssim 0.95
\label{nnh}
\end{equation}
which can be conveniently used, e.g, to perform magnetoresistance experiments. An example of the voltage-phase dependence obtained in this region is presented in Fig. \ref{mr1}. These plots demonstrate {\it negative magnetoresistance}, i.e. the system resistance decreases with increasing magnetic flux $\Phi$. The amplitude of this voltage modulation effect decreases with increasing dephasing parameter $\gamma$.  Also for the values of $\chi$ sufficiently close to $\pi$ the hysteretic behavior can be reinstated. E.g. for plots displayed in  Fig. \ref{mr1} this is the case at $\chi\gtrsim 2.5$.

\begin{figure}
\includegraphics[width=7.8cm]{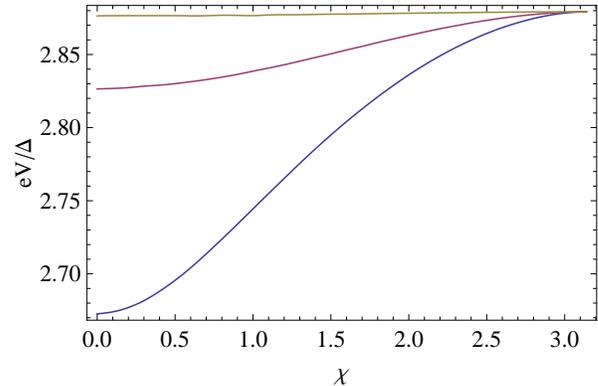}
\caption{ The same as in Fig. 6 at higher bias current value $I= 2.7\Delta/(e R_N)$. The values of $\gamma$ are 0.1, 0.8, 4 (bottom to top). }
\label{mr2}
\end{figure}

Note that the effect of voltage modulation by the external flux persists also at higher voltages $eV>\Delta+\Delta_g$, in which case deviations from the normal state behavior become smaller. In this regime the system magnetoresistance turns out to be positive, though the amplitude of voltage modulations is diminished and becomes vanishingly small for sufficiently large values of $\gamma$. This behavior is exemplified in Fig. \ref{mr2}.

\section{Discussion}
In this work we analyzed the behavior of Andreev interferometer in the current-biased regime. As compared to the voltage-biased regime studied before \cite{GZK}, we discover a rather different form of the $I-V$ curves and observe some peculiarities, e.g., at voltage values defined in Eqs. (\ref{p1}) and (\ref{p2}). Singularities at voltages (\ref{p1}) are due to the Riedel-like feature and are qualitatively similar to those observed in experiments with ordinary Josephson junctions \cite{Niem}. Additional features at voltages (\ref{p2}) are specific to systems under consideration and are not present, e.g., in tunnel junctions between two BCS superconductors.

An experimental observation of the above two series of singularities in Andreev interferometers under consideration would be highly interesting
as, for instance, it would allow to accurately determine the parameters of such devices. Indeed, the dephasing parameter $\gamma$ can be determined with the aid of Eq. (\ref{mge}), e.g., by setting an external magnetic flux equal to zero $\Phi =0$ (and, hence, $\epsilon_g=\Delta$). Having established this parameter one can also
verify the flux dependence $\epsilon_g (\Phi )$ (\ref{bex})  at all
values of $\Phi$ and also obtain extra information about the conductances $G_2$ and $G_3$.

Depending on the value of the applied external current Andreev interferometers can show either hysteretic or non-hysteretic behavior.
The latter behavior is well suited to study the effect of voltage modulation by an external magnetic flux. E.g. within the voltage interval (\ref{nnh}) one observes negative magnetoresistance and significant voltage modulation effect (cf. Fig. 6) which is sufficient for reliable performance of Andreev interferometers.

Let us also note that the above features of Andreev interferometers predicted here, such as hysteretic behavior, peaks in the differential resistance and negative magnetoresistance have been observed in recent
experiments \cite{Petrnew}. It appears, however, that more work will be needed in order to perform a quantitative comparison between theory and experiment.

To complete our discussion we briefly address the effect of voltage noise.
Denoting the voltage fluctuation by  $\delta V(t)$, introducing the voltage-voltage correlation function
\begin{equation}
S_V(\omega)=\int d(t_1-t_2)e^{i\omega(t_1-t_2)}\left\langle \delta V(t_1)\delta V(t_2)\right\rangle.
\end{equation}
and following the analysis developed in Ref. \onlinecite{Zor} for the case of ordinary Josephson junctions (see also Ref. \onlinecite{Likh}), in the limit of low frequencies for $eV\lesssim 2\Delta$ we arrive at an estimate
 \begin{equation}
 S_V(0)\sim \Delta R_d^2/R_N, \quad R_d=dV/dI.
 \label{estn}
 \end{equation}
 We note  that within voltage interval (\ref{nnh}) the differential resistance of our device obeys the inequality $R_d\lesssim R_N$, see Figs. 3-5. Since the voltage modulation for small $\gamma$ is $\sim \Delta/e$, see Fig. 6,  a typical Noise-to-Signal-Ratio (NSR) of our device can be estimated as
 \begin{equation}
 {\rm NSR}\lesssim \sqrt{(R_N/R_q)(\delta\omega/\Delta )}. \label{estn2}
 \end{equation}
 Here $R_q$ is the quantum resistance unit and $\delta\omega$ defines the bandwidth for our system. As here we are interested in the averaged voltage and as higher Josephson harmonics should be effectively filtered out,  we may set $\delta\omega\ll \Delta$.  In addition, we will assume that $R_N \ll R_q$. In this case the estimate (\ref{estn2}) yields ${\rm NSR} \ll 1$. Perhaps, we may also add that, as it was concluded in recent experiments \cite{MM,Petrnew}, intrinsic noise of Andreev interferometers was lower than that in employed readout electronics. This observation combined with the estimate (\ref{estn2}) appears to indicate that voltage noise may remain sufficiently weak and will not compromise the performance of Andreev interferometers.

\vspace{0.3cm}

\centerline{\bf Acknowledgements}

We appreciate valuable discussions with V.T. Petrashov and J. Pekola.
The work was supported by the Act 220 of the Russian Government (project 25).

\appendix
\section{}
In the case of a tunnel barrier between two BCS superconductors the expressions for $S_{1,2}$ (\ref{gr}) reduce to the well known results \cite{W,LO} which can be briefly summarized as follows.

Following Werthamer \cite{W} let us introduce the notations
\begin{equation}
x=\frac{|\omega|}{\Delta_1+\Delta_2}, \quad \delta= \frac{|\Delta_1-\Delta_2|}{\Delta_1+\Delta_2}.
\end{equation}
Then at $T \to 0$ we find
\begin{widetext}
\begin{eqnarray}
&&\tilde S_1(\omega)=\frac{2\Delta_1\Delta_2}{eR_N(\Delta_1+\Delta_2)} \left( \frac{1}{\sqrt{1-x^2}}K\left( \frac{\delta^2-x^2}{1-x^2}\right)-\frac{2\sqrt{1-x^2}}{1-\delta^2}E\left( \frac{\delta^2-x^2}{1-x^2}\right)\right),\quad 0\le x\le \delta; \nonumber
\\ && \tilde S_1(\omega)=\frac{2\Delta_1\Delta_2}{eR_N(\Delta_1+\Delta_2)} \frac{1}{\sqrt{1-\delta^2}}\left( K\left( \frac{x^2-\delta^2}{1-\delta^2}\right)-2 E\left( \frac{x^2-\delta^2}{1-\delta^2}\right)\right),\quad \delta\le x\le 1; \label{S1}
\\ && \tilde S_1(\omega)=\frac{2\Delta_1\Delta_2}{eR_N(\Delta_1+\Delta_2)} \left[ \frac{2\sqrt{x^2-\delta^2}}{1-\delta^2}\left( K\left( \frac{1-\delta^2}{x^2-\delta^2}\right)-E\left( \frac{1-\delta^2}{x^2-\delta^2}\right)\right) -\frac{1}{ \sqrt{x^2-\delta^2}}K\left( \frac{1-\delta^2}{x^2-\delta^2}\right)\right. \nonumber
\\ &&\left. +i\, {\rm sgn}\,\omega\left( \frac{2\sqrt{x^2-\delta^2}}{1-\delta^2}E\left( \frac{x^2-1}{x^2-\delta^2}\right) -\frac{1}{\sqrt{x^2-\delta^2}}K\left( \frac{x^2-1}{x^2-\delta^2}\right) \right)\right],\quad x\ge 1\nonumber
\end{eqnarray}
and
\begin{eqnarray}
&& \tilde S_2(\omega)=\frac{2\Delta_1\Delta_2}{eR_N(\Delta_1+\Delta_2)} \frac{1}{\sqrt{1-x^2}}K\left( \frac{\delta^2-x^2}{1-x^2}\right),\quad 0\le x\le \delta; \nonumber
\\  &&\tilde S_2(\omega)=\frac{2\Delta_1\Delta_2}{eR_N(\Delta_1+\Delta_2)} \frac{1}{\sqrt{1-\delta^2}}K\left( \frac{x^2-\delta^2}{1-\delta^2}\right),\quad\delta\le x\le 1;\label{S2}
\\  && \tilde S_2(\omega)=\frac{2\Delta_1\Delta_2}{eR_N(\Delta_1+\Delta_2)} \frac{1}{\sqrt{x^2-\delta^2}}\left(  K\left( \frac{1-\delta^2}{x^2-\delta^2}\right)+ i\, {\rm sgn}\,\omega K\left( \frac{x^2-1}{x^2-\delta^2}\right) \right),\quad x\ge 1, \nonumber
\end{eqnarray}
\end{widetext}
where
\begin{equation}
K(k)=\int\limits_0^{\pi/2} \frac{d\phi}{\sqrt{1-k\sin^2\phi}},\; E(k)=\int\limits_0^{\pi/2} d\phi\sqrt{1-k\sin^2\phi}
\end{equation}
are complete elliptic integrals.
At $x\to 1$ these expressions diverge demonstrating the so-called Riedel singularity \cite{Ried}
\begin{eqnarray}
&& {\rm Re}\,\tilde S_1(\omega)\approx \frac{\sqrt{\Delta_1\Delta_2}}{2eR_N}\left[ \ln\left( \frac{8(1-\delta^2)}{|1-x|}\right)-4\right],\nonumber\\
&& {\rm Re}\,\tilde S_2(\omega)\approx \frac{\sqrt{\Delta_1\Delta_2}}{2eR_N}\ln\left( \frac{8(1-\delta^2)}{|1-x|}\right). \label{rs}
\end{eqnarray}


\begin{thebibliography}{99}
\bibitem{Petr1} V.T. Petrashov, V.N. Antonov, P. Delsing, and T. Claeson,
 Phys. Rev. Lett. {\bf 70}, 347 (1993); JETP Lett. {\bf 60}, 589 (1994);
 Phys. Rev. Lett. {\bf 74}, 5268 (1995).
 \bibitem{Petrreadout} V.T. Petrashov, K.G. Chua, K.M. Marshall, R.Sh. Shai\-khaidarov, and J.T. Nicholls, Phys. Rev. Lett. {\bf 95}, 147001 (2005).
 \bibitem{Petrdynam} I. Sosnin, H. Cho, V.T. Petrashov, and A.F. Volkov,  Phys. Rev. Lett. {\bf 96}, 157002 (2006).
\bibitem{MM} M. Meschke, J.T. Peltonen, J.P. Pekola, and F. Giazotto,
    Phys. Rev. B {\bf 84}, 214514 (2011).
 \bibitem{GT} F. Giazotto and F. Taddei, Phys. Rev. B {\bf 84}, 214502 (2011).
 \bibitem{GZK} A.V. Galaktionov, A.D. Zaikin, and L.S. Kuzmin, Phys. Rev. B {\bf 85},  224523 (2012).
\bibitem{Reul} B. Reulet, J. Senzier, and D.E. Prober, Phys. Rev. Lett. {\bf 91}, 196601 (2003).
\bibitem{Levitov2} L.S. Levitov and M. Reznikov, Phys. Rev. B \textbf{70}, 115305 (2004).
\bibitem{3rdcum} A.V. Galaktionov, D.S. Golubev, and A.D. Zaikin, Phys. Rev. B {\bf 68}, 235333 (2003).
\bibitem{Salo} J. Salo, F.W.J. Hekking, and J. Pekola, Phys. Rev. B {\bf 74}, 125427 (2006).
\bibitem{Kind} M. Kindermann, Yu.V. Nazarov, and C.W.J. Beenakker,  Phys. Rev. B {\bf 69}, 035336 (2004).
\bibitem{McD} D.G. McDonald, E.G. Johnson, and R.E. Harris, Phys. Rev. B {\bf 13}, 1028 (1976).
\bibitem{Likh} K.K. Likharev, {\it Dynamics of Josephson Junctions and Cir\-cuits} (Gordon and Breach Science Publishers, Amsterdam, 1986).
\bibitem{Niem} J. Niemeyer and V. Kose, Appl. Phys. Lett. {\bf 29}, 380 (1976).
\bibitem{BFSZ} P. Bobbert, R. Fazio, G. Sch\"{o}n, and A.D. Zaikin, Phys. Rev. B {\bf 45}, 2294 (1992).
\bibitem{GWZ} A.A. Golubov, F.K. Wilhelm, and A.D. Zaikin, Phys. Rev. B
    {\bf 55}, 1123 (1997).
\bibitem{NazSt} Yu.V. Nazarov and T.H. Stoof, Phys. Rev. Lett. {\bf 76}, 823 (1996).
\bibitem{SZ} G. Sch\"on and A.D. Zaikin, Phys. Rep. {\bf 198}, 237 (1990).
\bibitem{Z94} A.D. Zaikin, Physica B {\bf 203}, 255 (1994).
\bibitem{Sny} I. Snyman and Yu.V. Nazarov, Phys. Rev. B {\bf 77}, 165118 (2008).
\bibitem{GZ10} A.V. Galaktionov and A.D. Zaikin, Phys. Rev. B {\bf 82}, 184520 (2010).
\bibitem{BSW} E.V. Bezuglyi, V.S. Shumeiko, and G. Wendin, Phys. Rev.
    B {\bf 68},    134506 (2003).
\bibitem{W} N.R. Werthamer, Phys. Rev. {\bf 147}, 255 (1966).
\bibitem{LO} A.I. Larkin and Yu.N. Ovchinnikov, Sov. Phys. JETP {\bf 24}, 1035 (1967).
\bibitem{avail} Our C++ numerical code is available upon request.
\bibitem{Ried} E. Riedel, Z. Naturforsch. {\bf 19A}, 1634 (1964).
\bibitem{Petrnew} J. Wells, C. Shelly, and V.T. Petrashov. {\it Superconducting Transitions and Quantum Interference in Hybrid Nanostructures}, Royal Holloway, University of London, Report (2013).
\bibitem{Zor} A.B. Zorin, Sov. J. Low Temp. Phys. {\bf 7}, 709 (1981).
    \end{thebibliography}
\end{document}